\documentclass[conference]{IEEEtran}
\IEEEoverridecommandlockouts
\usepackage{cite}
\usepackage{amsmath,amssymb,amsfonts}
\usepackage{algorithmic}
\usepackage{graphicx}
\usepackage{textcomp}
\usepackage{xcolor}
\usepackage{multirow}
\usepackage{graphicx}
\usepackage{xcolor}
\usepackage{soul}
\def\BibTeX{{\rm B\kern-.05em{\sc i\kern-.025em b}\kern-.08em
    T\kern-.1667em\lower.7ex\hbox{E}\kern-.125emX}}

\begin{document}

\title{Rebuild AC Power Flow Models with Graph Attention Networks}
\author{
  Yuting Hu \quad
  Jinjun Xiong \\
  University at Buffalo, Buffalo, NY, USA \\
  \texttt{\{yhu54, jinjun\}@buffalo.edu} \\
}

\maketitle

\begin{abstract}
A full power flow (PF) model is a complete representation of the physical power network. Traditional model-based methods rely on the full PF model to implement power flow analysis. In practice, however, some PF model parameters can be inaccurate or even unavailable due to the uncertainties or dynamics in the power systems. Moreover, because the power network keeps evolving with possibly changing topology, the generalizability of a PF model to different network sizes and typologies should be considered. In this paper, we propose a PF rebuild model based on graph attention networks (GAT) by constructing a new
graph based on the real and imaginary parts of voltage at each bus.
By comparing with two state-of-the-art PF rebuild models for different standard IEEE power system cases and their modified topology variants, we demonstrate the feasibility of our method. Experimental results show that, 
our proposed model achieves better accuracy for a changing network and can
generalize to different networks with less accuracy discount. 
\end{abstract}


\section{Introduction}
\noindent
Power flow (PF) analysis is a  foundational tool for power system analysis, including its operations, planning, monitoring, security assessment, and state analysis [1]. It tries to obtain the complete voltage angles and magnitudes at all buses in the power system by solving a series of PF equations. For accurate analysis, AC power flow has to be determined since it fully describes the physical power system by following the Kirchoff's laws, satisfying system operational constraints, and incorporating the power network's topology and the corresponding parameters, which is called a \textit{full PF model} or a \textit{PF configuration} [2]. Traditionally, the mainstream method for solving ACPF is based on the Newton-Raphson (NR) method, which requires the knowledge of the admittance matrix of the power system to formulate the Jacobian matrix at every iteration step [3]. 
Thus the NR-based ACPF is model-based and needs the complete knowledge of the full PF model.

However, with the increasing uncertainty and dynamics in modern power systems with renewable integration, distributed energy resources, and EV charging and storage, the system parameters required by PF models can be inaccurate or even unavailable. To be specific, the PF models are usually partially observable due to line aging, fluctuating weather conditions, and missing information in secondary distribution grids [4]. Therefore, recent research on power flow modeling focuses on rebuilding PF models by identifying system parameters via data-driven methods, i.e, building the mapping rules between voltage phasors and power phasors. Since a rebuilt PF model can capture the constantly changing PF systems, it can be used to initialize state estimators, recorrect PF solvers, calibrate PF models, and ensure optimization convergence [5]. 

To build the mapping rules of PF models, some data-driven machine learning methods have been proposed such as the linear regression based on partial least squares (PLS-LR) [6], Bayesian linear regression (BLR) [6], and support vector regression (SVR) [7]. Besides, deep neural network models (e.g., Multi-layer Perceptron (MLP) [8], Bilinear Neural Network [9]) also exhibit high degree of accuracy to approximate PF models. However, all these methods aim at inferring the mapping rules of PF models for a given power network topology by learning from the network's historical measurement data. They cannot be applied to different network typologies, which limits their applications only to the same power network with different parameters.

Since power grids can be naturally represented as a graph data structure, it has generated much research interest recently to apply graph neural networks (GNNs) [10] to solve power grid tasks. GNNs are neural models for the graph-structured data. It embeds the graph-structured data into a low-dimension vector space while preserving necessary information for some downstream machine learning tasks such as link prediction [11] and node classification [12]. The two widely used GNN models are graph convolutional networks (GCN) [13] and graph attention networks (GAT) [14]. Different from GCN in which the layer propagation is fully determined by the graph structure, GAT learns to reweight different node features in a neighborhood so that it can better support both inductive and transductive problems. By modeling the buses and power lines as the nodes and edges respectively, and representing the network parameters as the node or edge features, a physical power network can be completely represented as a graph. A few recent novel works [15-18] propose to use GNNs to indirectly solve the ACPF problems. Different graph construction methods are proposed in their work. However, if a power network's parameters are not completely known, we cannot build such graphs for power grids and all existing methods [15-18] and other full PF model-based methods will become infeasible. Therefor, a new PF rebuild model is needed to consider different power networks for their PF solver.

The contributions of this paper can be summarized as follows:
\begin{itemize}
\item 
Under the assumption that a power network topology is known but its network parameters can be inaccurate or unavailable, 
we propose a novel method to rebuild the PF  model based on graph attention neural networks (GAT).
\item
The proposed GAT model can rebuild a PF model for the ACPF analysis of different power networks with varying topologies and parameters.
\item We compare our GAT model with two state-of-the-art rebuilding PF models on different standard IEEE cases,
and experimental results show that the proposed GAT model can achieve comparable accuracy and much more improved generalization capabilities when applied to different power networks. 
\end{itemize}

This remaining paper is structured as follows. Section~\ref{sec:model} 
reviews the ACPF analysis and describes the architecture of our GAT-based mapping model. Section~\ref{sec:exp} 
evaluates the accuracy and generalizability of the proposed GAT model on different IEEE power system cases,
and analyzes experimental results to illustrate the advantages of our proposed GAT model. Section~\ref{sec:diss} 
concludes the paper with discussions of our future work.

\section{Our GAT Model for Rebuilding PF Models}
\label{sec:model}

\subsection{The AC Power Flow Problem and Graph Construction}
\noindent
Based on the complex power flow analysis, 
the conventional ACPF problem
can be formulated as a system of nonlinear equations as follows:
\begin{equation}
S_{i} = U_{i}I_{i}^{*} = U_i\left ( \sum_{j=1}^{n}Y_{ij}U_{j} \right )^{*},
\label{eqn:acpf}
\end{equation}
\noindent
where $S_{i}$, $U_{i}$ and $I_i$ are the complex power injection, voltage phasors, and current injection at bus $i$, respectively; and $Y_{i,j}$ is the (i,j) element in the admittance matrix $Y$. 
In the Cartesian form, we have $S_{i} = P_{i} + jQ_{i}$, where $P_{i}$ and $Q_{i}$ are the active and reactive power injections, respectively.

The objective of power flow analysis is to obtain the complex voltages $U=\mu+j\omega$ at all buses. We denote the voltage vector as follows:
$\mu = [\mu_{1},\mu_{2},...,\mu_{n}]$, $\omega = [\omega_{1},\omega_{2},...,\omega_{n}]$, and $U=[\mu,\omega]$.
The PF equations is based on minimizing the power mismatches at all buses as follows:
\begin{equation}
f([\mu,\omega])=\begin{bmatrix}
P'-P([\mu,\omega ])\\ 
Q'-Q([\mu,\omega ])
\end{bmatrix}
=0
\end{equation}
where  $P'$ and $Q'$ are known active and reactive power injections, respectively. 


In the Cartesian form,  $Y_{i,j}=G_{i,j}+jB_{i,j}$ with $G_{ik}$ and ${B_{ik}}$ as the real and imaginary parts $Y_{i,j}$.
Therefore, the ACPF problem can be further expressed as follows:
\begin{equation}
P_{i}=\sum_{k=1}^{n}G_{ik}(\mu_{i}\mu_{k}+\omega_{i}\omega_{k})+B_{ik}(\omega_{i}\mu_{k}-\mu_{i}\omega_{k})
\end{equation}
\begin{equation}
Q_{i}=\sum_{k=1}^{n}G_{ik}(\omega_{i}\mu_{k}-\mu_{i}\omega_{k})-B_{ik}(\mu_{i}\mu_{k}+\omega_{i}\omega_{k})
\end{equation}
Since the power injections are linear combinations of the $2^{th}$ Kronecker power of bus voltages, the SVR model [7], bilinear neural network (BNN) model [9], and topology-pruned BNN model [9] are then proposed to rebuild the PF models. 

In our setting, however, the admittance matrix is unknown. To rebuild the PF model, we propose to use a data-driven method to learning mapping rules from the data pairs of input and output, which can better capture the changes of the PF models. 
%
%
We note that the conjugated current injection 
into bus $i$, i.e., $I_{i}^{*}=I_{ri}^{*}+jI_{ii}^{*}$,
can be expressed in the form of separate real and imaginary parts as follows: 
\begin{equation}
\begin{split}
\scriptsize{
\setlength{\arraycolsep}{0.4pt}
\begin{bmatrix}
I_{r1}^{*}\\ 
I_{r2}^{*}\\ 
\vdots \\ 
I_{rn}^{*}\\ 
I_{i1}^{*}\\ 
I_{i2}^{*}\\ 
\vdots\\ 
I_{in}^{*}
\end{bmatrix}=\begin{bmatrix}
&G_{1,1} &G_{1,2}  &\cdots   &G_{1,n}  &-B_{1,1}  &-B_{1,2}  &\cdots  &-B_{1,n} \\ 
&G_{2,1} &G_{2,2}  &\cdots   &G_{2,n}  &-B_{2,1}  &-B_{2,2}  &\cdots  &-B_{2,n} \\ 
&\vdots  &\vdots   &\ddots   &\vdots   &\vdots  &\vdots  &\ddots  &\vdots \\ 
&G_{n,1} &G_{n,2}  &\cdots   &G_{n,n}  &-B_{n,1}  &-B_{n,2}  &\cdots  &-B_{n,n} \\ 
&-B_{1,1}  &-B_{1,2}  &\cdots  &-B_{1,n} &-G_{1,1} &-G_{1,2}  &\cdots   &-G_{1,n} \\ 
&-B_{2,1}  &-B_{2,2}  &\cdots  &-B_{2,n} &-G_{2,1} &-G_{2,2}  &\cdots   &-G_{2,n} \\ 
&\vdots    &\vdots    &\ddots  &\vdots   &\vdots  &\vdots   &\ddots   &\vdots \\ 
&-B_{n,1}  &-B_{n,2}  &\cdots  &-B_{n,n} &-G_{n,1} &-G_{n,2}  &\cdots   &-G_{n,n}
\end{bmatrix}
\begin{bmatrix}
\mu_{1}\\ 
\mu_{2}\\ 
\vdots \\ 
\mu_{n}\\ 
\omega_{1}\\ 
\omega_{2}\\ 
\vdots\\ 
\omega_{n}
\end{bmatrix}
}
\end{split}
\label{eqn:conjugate_I}
\end{equation}

\begin{figure}[htbp]
\centerline{\includegraphics[width=9cm]{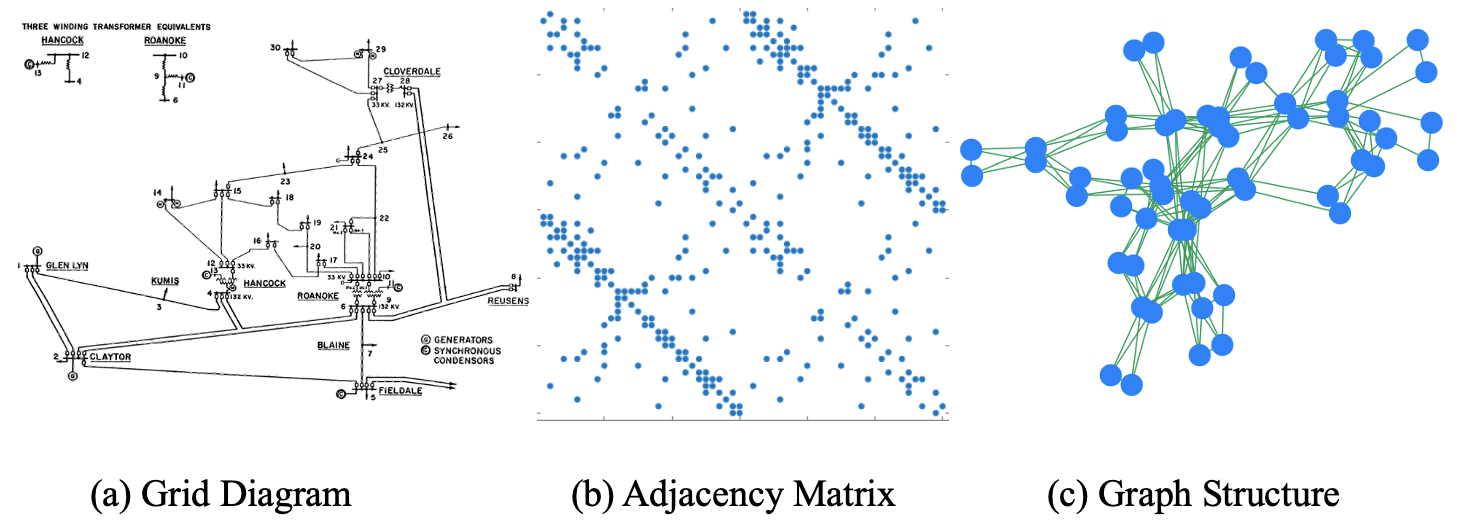}}
\caption{Construction of the Proposed Graph for a Known Network Topology.}
\label{fig:gnn}
\end{figure}

Our graph construction is based on the perspective of current computing as shown in (\ref{eqn:conjugate_I}).
In our formulation, we model $\mu$ and $\omega$ as nodes in the graph, which is significantly different from other GNN-based methods where  power lines or buses are represented as graph nodes. 
Assuming that the power system topology is available, we can encode the network structure information into the graph as follows. 
Since the admittance matrix $Y$ reflects the connection relationships between buses for a given network topology, we can easily get an adjacency matrix $A\in  R^{2n\times 2n}$ to construct the corresponding graph. As shown in Figure~\ref{fig:gnn}, the constructed graph contains $2n$ nodes with every node containing one attribute representing the value of either $\mu$ or $\omega$. 

 In the context of GNN, the current calculation in (\ref{eqn:conjugate_I}) can be treated as an update of node attributes after the weighted aggregation of neighbor node features. The weighted matrix formed by power network admittance is what we need to learn for the GNN-based PF model.
 
\subsection{Data-driven GAT-based PF Model}
\noindent
For graph neural networks (GNNs), with one layer, the input is a set of node features which can be represented by $\mathbf{h} = \left \{ h_{1},h_{2},\cdots,h_{N}\right \}, h_{i}\in R^{d}$, where $N$ is the number of nodes and $d$ is the number of features at each node. The output is an updated set of node features $\mathbf{h'} = \left \{ h'_{1},h'_{2},\cdots,h'_{N}\right \}, h'_{i}\in R^{d'}$. In our case, $d'$ is the number of channels in the hidden layers, while the number of nodes $N$ is twice the size of the bus numbers, i.e., $N=2n$. And the weighted matrix $W\in R^{d'\times d}$ is applied to every node. 
\begin{figure}[htbp]
\centerline{\includegraphics[width=6cm]{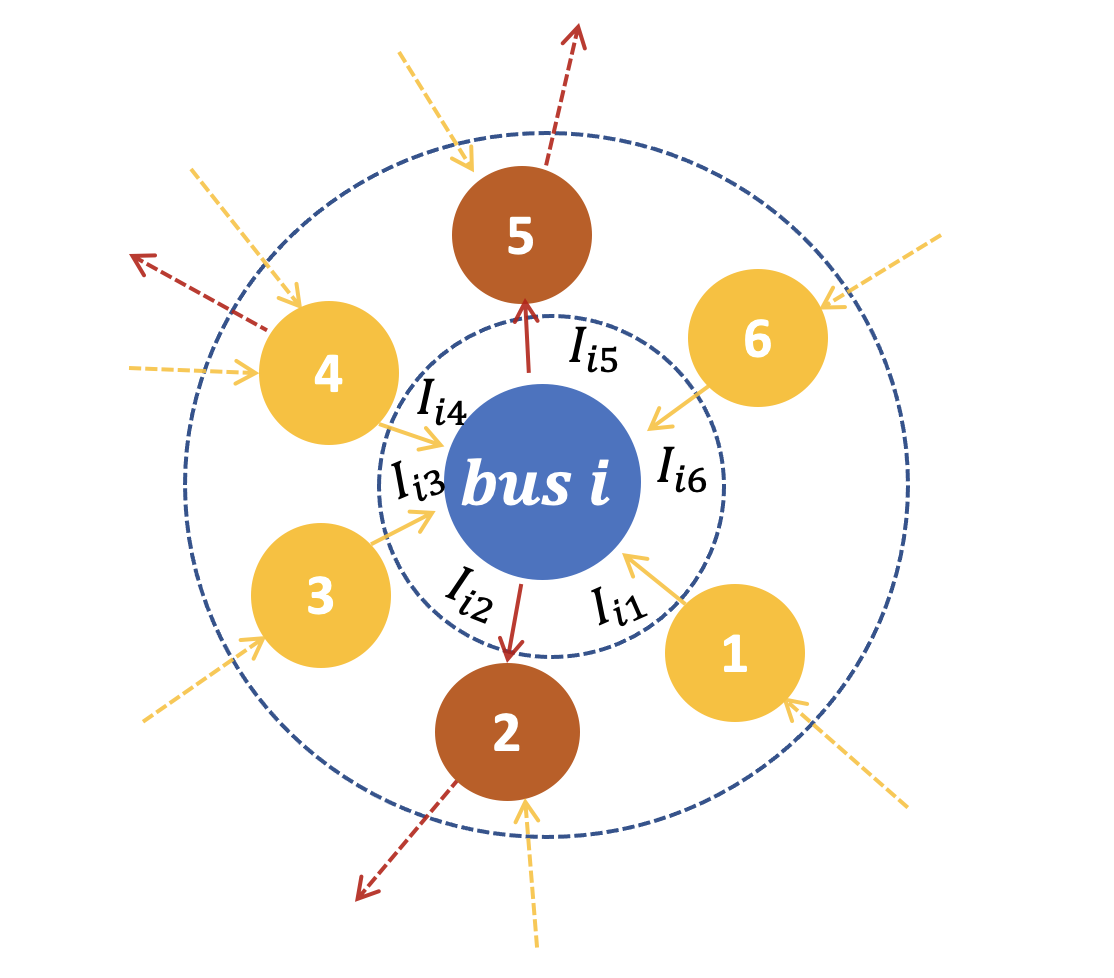}}
\caption{Attention-like Current Injections at Power Grid Buses.}
\label{fig:attention}
\end{figure}

In a graph attention network (GAT), the attention mechanism is a feed-forward neural network by applying LeakyRelu activation. The network parameters are represented by $\mathbf{a}\in R^{2d'}$
In GAT, the attention coefficients $\alpha _{ij}$ is computed as follows:
\begin{equation}
\alpha _{ij}=\frac{exp(\textup{LeakyRelu}(\mathbf{a}^{T}[Wh_{i}||Wh_{j}]))}{\sum_{k\in N_{i}\cup\left \{ i \right \}}exp(\textup{LeakyRelu}(\mathbf{a}^{T}[Wh_{i}||Wh_{k}])) }
\end{equation}
where $N_i$ is the neighboring nodes of node $i$, the $||$ is the concatenate operation. The attention coefficients $\alpha _{ij}$ intuitively indicate the importance of node $j$'s features to node $i$. In a power network, the larger the mutual admittance between two nodes, the easier it is for current to flow between them. Thus the current injection at a bus is decided by its voltage and the voltage of nodes in its neighborhood,  its self-admittance, and mutual admittance along the connected edges. As shown in Figure~\ref{fig:attention}, 
attention mechanisms enable different weights to different nodes in a neighborhood. 
With attention coefficients, the neighboring features around bus $i$ can be linearly aggregated and passed forward to the next layer. After applying activation function, the output features of bus $i$ will be:
\begin{equation}
h'_{i} = \sigma (\sum_{j\in N_{i}}\alpha _{ij}h_{j}+\alpha _{ii}h_{i})
\end{equation}

\begin{figure}[htbp]
\centerline{\includegraphics[width=7cm]{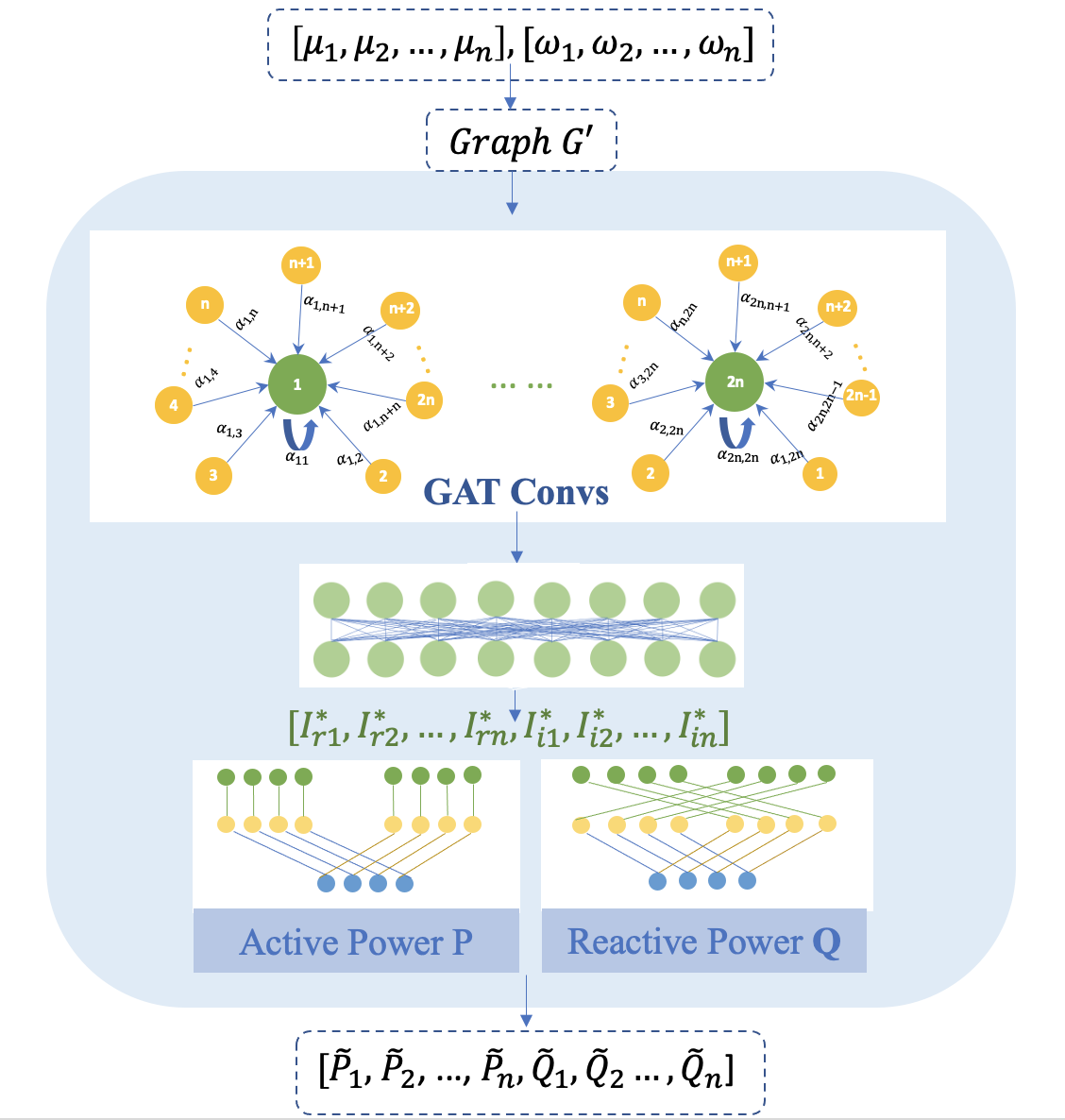}}
\caption{GAT-based Model Structure for Rebuilding PF Models.}
\label{fig:GAT4PF}
\end{figure}

We now describe how we rebuild PF models with incorporated typology information. 
We aim at providing a data-driven PF model that can adapt to various power networks with different sizes and topology configurations after sufficient training. 
To learn the mapping rules from bus voltages to power injections, our model  builds the maps between the voltages and current injections by employing a GAT network, which can be seen as a black box with unknown bus admittance $G$ and $B$.
After GAT training, we can obtain current injection $I_{i}^*$ from voltage phasors at all buses.
Per (\ref{eqn:acpf}), the power injection can thus be calculated. The proposed method to rebuild the PF model can be depicted in Figure~\ref{fig:GAT4PF}, which intuitively indicates the flow of our proposed GAT-based PF model.
Besides, as shown in (\ref{eqn:conjugate_I}), the weighted matrix formed by power network admittance has redundant parameters across its four sub-matrices. Thus, we can share the model weights between $\mu$ and $\omega$ nodes to reduce the number of learning parameters. 

\begin{table*}[tbp]
\caption{Comparison of RMSE for rebuilt PF under different models}
\label{tab:accuracy}
\vspace{-0.2in}
\begin{center}
\resizebox{\textwidth}{12mm}{
\begin{tabular}{|c|c||c|c|c|c|}
\hline
\multicolumn{2}{|c||}{\textbf{Base Cases}}  & \textbf{MLP$[8]$}& \textbf{TPBNN$[9]$}& \textbf{GAT*} & \textbf{GAT} \\
\hline
\multirow{2}{*}{IEEE 30} & $P$ & $2.63 \times 10^{-2}\pm 6.97 \times 10^{-4}$ & $6.93 \times 10^{-3}\pm 1.54 \times 10^{-4}$ & $\mathbf{3.83 \times 10^{-3}\pm 8.47 \times 10^{-4}}$ & $\mathbf{2.19 \times 10^{-2}\pm 1.93 \times 10^{-4}}$ \\
\cline{2-6} 
& $Q$ & $3.17 \times 10^{-2}\pm 8.22 \times 10^{-4}$ & $6.56 \times 10^{-3}\pm 4.37 \times 10^{-4}$ & $\mathbf{5.71 \times 10^{-3}\pm 3.92 \times 10^{-4}}$ & $\mathbf{3.61 \times 10^{-2}\pm 6.58 \times 10^{-4}}$ \\
\hline
\multirow{2}{*}{IEEE 57} & $P$ & $4.71 \times 10^{-2}\pm 8.95 \times 10^{-4}$ & $1.03 \times 10^{-2}\pm 6.81 \times 10^{-4}$ & $\mathbf{6.14 \times 10^{-3}\pm 4.35 \times 10^{-4}}$ & $\mathbf{2.51 \times 10^{-2}\pm 3.68 \times 10^{-4}}$ \\
\cline{2-6} 
& $Q$ & $3.64 \times 10^{-2}\pm 9.12 \times 10^{-4}$ & $8.39 \times 10^{-3}\pm 1.36 \times 10^{-4}$ & $\mathbf{9.83 \times 10^{-3}\pm 3.26 \times 10^{-4}}$ & $\mathbf{2.92 \times 10^{-2}\pm 1.73 \times 10^{-4}}$ \\
\hline
\multirow{2}{*}{IEEE 118} & $P$ & $3.98 \times 10^{-2}\pm 1.67 \times 10^{-4}$ & $1.94 \times 10^{-2}\pm 5.27 \times 10^{-4}$ & $\mathbf{1.17 \times 10^{-2}\pm 6.49 \times 10^{-4}}$& $\mathbf{2.86 \times 10^{-2}\pm 6.72 \times 10^{-4}}$ \\
\cline{2-6} 
& $Q$ & $4.21 \times 10^{-2}\pm 6.55 \times 10^{-4}$ & $7.83 \times 10^{-3}\pm 4.92 \times 10^{-4}$ & $\mathbf{1.34 \times 10^{-2}\pm 6.51 \times 10^{-4}}$& $\mathbf{3.27 \times 10^{-2}\pm 4.58 \times 10^{-4}}$\\
\hline
\end{tabular}
}
\end{center}
\end{table*}

\begin{table*}[tbp]
\caption{Comparison of RMSE for PF solver under different models}
\label{tab:accuracy}
\vspace{-0.2in}
\begin{center}
\resizebox{\textwidth}{12mm}{
\begin{tabular}{|c|c||c|c|c|}
\hline
\multicolumn{2}{|c||}{\textbf{Base Cases}}  & \textbf{MLP}& \textbf{GAT+MLP}& \textbf{GAT+MLP+Regularizer} \\
\hline
\multirow{2}{*}{IEEE 30} & $\mu$ & $2.45 \times 10^{-2}\pm 2.47 \times 10^{-4}$ & $4.38 \times 10^{-3}\pm 1.29 \times 10^{-4}$ & $\mathbf{1.22 \times 10^{-3}\pm 6.43 \times 10^{-4}}$ \\
\cline{2-5} 
& $\omega$ & $2.71 \times 10^{-2}\pm 6.28 \times 10^{-4}$ & $5.51 \times 10^{-3}\pm 6.23 \times 10^{-4}$ & $\mathbf{1.83 \times 10^{-3}\pm 4.58 \times 10^{-4}}$ \\
\hline
\multirow{2}{*}{IEEE 57} & $\mu$ & $4.42 \times 10^{-2}\pm 3.93 \times 10^{-4}$ & $1.14 \times 10^{-2}\pm 6.52 \times 10^{-4}$ & $\mathbf{2.38 \times 10^{-3}\pm 2.15 \times 10^{-4}}$ \\
\cline{2-5} 
& $\omega$ & $3.65 \times 10^{-2}\pm 8.25 \times 10^{-4}$ & $9.58 \times 10^{-3}\pm 4.22 \times 10^{-4}$ & $\mathbf{2.76 \times 10^{-3}\pm 3.71 \times 10^{-4}}$ \\
\hline
\multirow{2}{*}{IEEE 118} & $\mu$ & $4.91 \times 10^{-2}\pm 3.64 \times 10^{-4}$ & $1.87 \times 10^{-2}\pm 4.36 \times 10^{-4}$ & $\mathbf{6.39 \times 10^{-2}\pm 6.12 \times 10^{-4}}$ \\
\cline{2-5} 
& $\omega$ & $4.38 \times 10^{-2}\pm 5.57 \times 10^{-4}$ & $2.39 \times 10^{-2}\pm 8.01 \times 10^{-4}$ & $\mathbf{7.04 \times 10^{-2}\pm 5.53 \times 10^{-4}}$\\
\hline
\end{tabular}
}
\end{center}
\end{table*}

\section{Experimental Results}
\noindent
In this section, we perform three experiments to test our model's validity and generalizability. In the first experiment, we test and compare our GAT-based PF rebuild model with MLP [8] and Typology-Pruned BNN (TPBNN) [9] on various standard IEEE bus cases. In the second experiment, for these models, we evaluate their generalization capability of adapting to unseen typologies. In the third experiment, we assess the generalization ability of our model to new power networks with different sizes and typologies that are not even available in the training process.

\subsection{Experiment Setup}\label{AA} 
\noindent 
For our experiment, we need the dataset composed by ACPF solutions of various power networks. Thus, We use the following topology sampling process to generate different power network cases. We first take an existing IEEE network case as a base case. Then for a base case with $n$ bus nodes and $m$ branches, we generate a number of variant cases, each of which will have the same number of $n$ bus nodes with the base case. This can be achieved by (1) generating a minimum spanning tree of the $n$ bus nodes from the base case, and (2) then randomly adding additional branches to the spanning tree until the total number of generated branches is $m$. (3) for the randomly generated branches, we randomly map them to the base case's branches; (4) for each mapped branch, its branch resistance, reactance and charging susceptance are uniformly sampled from [90\%, 110\%] of their corresponding base branches' reference values. 

To generate different power injections for each case to get ACPF solution instances as dataset, we use the following process:(1) we keep the node types (PV or PQ bus) as same as the base case; (2) for the PV bus node, its active power generation $P$ is uniformly sampled from the base node's reference value in the range of [75\%, 125\%], and its generator voltage magnitude $V$ (p.u.) is uniformly sampled from the base node's reference value in the range of [0.95, 1.05]; (3) for the PQ bus node, its active and reactive power demands are uniformly sampled from [50\%, 150\%] of their base $P$ and $Q$ values;  

By randomly sampling branches and adding noise on both injections and line parameters, we can acquire a variety of instances across different typologies. Then, we use MATPOWER to obtain the corresponding ACPF solutions of each instance. And only the instances that can be successfully solved by Newton Rapshon method in MATPOWER are included in the dataset. 

We implement our GAT model in PyTorch-Geometric [19]. For each experiment, we train all our GAT models with the Adam optimizer. The training/validation/test data are split as 60\%/10\%/30\% . We report the test results in next section.

\subsection{Evaluation Results with Benchmarks}
In this section, we evaluate the effectiveness of the proposed GAT-based model by comparing it with MLP [8] and Typology-pruned BNN [9] in PF rebuild tasks. For this experiment, we first generate dataset as shown in  TABLE~\ref{tab:dataset} for several standard IEEE cases by randomly sampling the power injections. Then, we separately train our model in the dataset for each case and in the combined dataset of all the cases in TABLE~\ref{tab:dataset} to make comparisons. Since the network sizes are different for each case, we pad zeros in the output of GAT to a fixed max length as shown in Figure 3. It needs to be specify that the MLPs and BNNs for each case are trained separately using the fixed network topology of each standard case since they cannot adapt to different network. However, our model can be trained only once with the combined dataset. 

\begin{table}[htbp]
\caption{Generated Dataset for Different Cases in Experiment 1}
\label{tab:dataset}
\vspace{-0.15in}
\begin{center}
\begin{tabular}{|c|c|c|c|}
\hline
\textbf{IEEE Base Cases} & \textbf{30 bus}& \textbf{57 bus}& \textbf{118 bus} \\
\hline
Branch Numbers& 41 & 80 & 186  \\
\hline
Data Set Sizes& 5000 & 8000 & 12,000  \\
\hline
\end{tabular}
\label{tab1}
\end{center}
\end{table}

TABLE ~\ref{tab:accuracy}
shows the average results in terms of root mean square error when evaluating our model with MLP and BNN in different cases. The {GAT*} column shows the RMSE results of separate training for each case. Compared to MLP and TPBNN, our {GAT*} models achieve smaller RMSEs. This convincingly shows the advantages of our GAT-based models. The {GAT} column shows the RMSE results of combined training for all the cases. Though the {GAT} model seems to produce higher error than the respective {GAT*} models, we note that their overall errors are still relative small and are better than MLP most of the time. 

Since the realworld power network is dynamical, even if MLP and TPBNN can achieve comparable accuracy at a fixed network, that doesn't mean they can maintain effective with changing networks. In the second experiment, we simulate the dynamic broken-branch scenarios in the power system and evaluate the performance of different models to adapt to unseen network topology. We use IEEE 57-bus as a base case and generate different variant cases with different typologies by gradually add random branch to its minimum spanning tree. For each variant case, we sample the power injections following the aforementioned process to get instances and solve them by MATPOWER to get the dataset. The dataset is composed by 120 variant cases of IEEE 57-bus with each variant case having 2000 ACPF solution instances. Then, we train our GAT model on the dataset. 

\begin{figure}[h]
\centerline{\includegraphics[width=7cm]{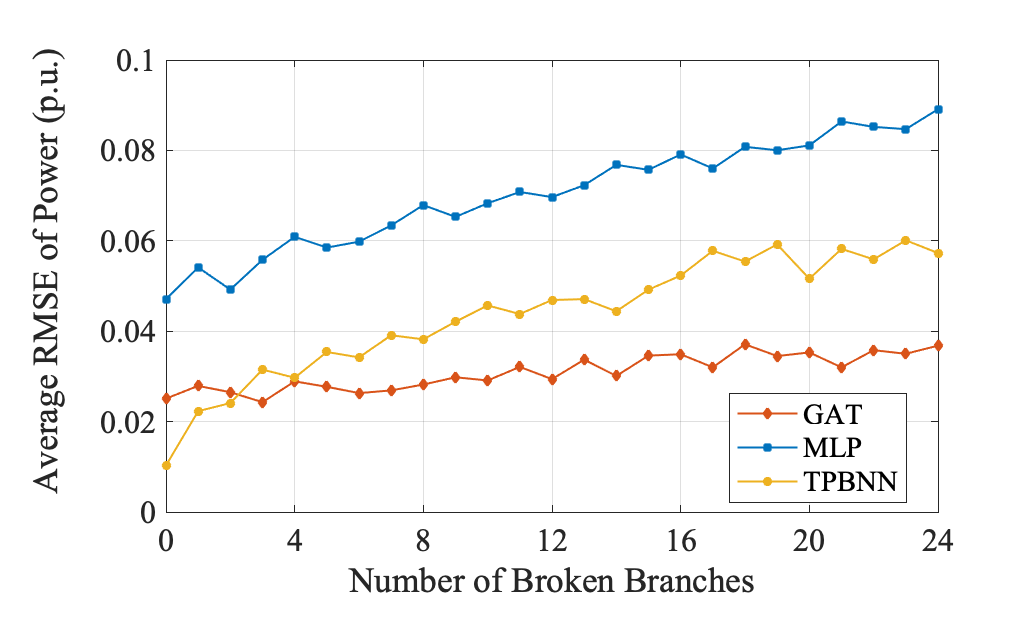}}
\caption{Rebuilt PF RMSE under varying IEEE 57-bus configurations.}
\label{fig:broken_branches}
\end{figure}
 
 For the test, we randomly remove some branches of IEEE 57-bus case. Since the IEEE 57-bus case has 80 branches, to keep all the 57 bus nodes fully connected (through a minimum spanning tree), at most 24 branches can be removed. We compare the average errors (i.e., RMSE) in the rebuilt PF solutions under different scenarios in Figure~\ref{fig:broken_branches}. It can be observed that our GAT model can generalize much better than the other two models for different topology scenarios, and when the scenarios become more challenging with more branches being removed, GAT's generalization errors stay relatively flat while the other two models' errors increase rapidly. This observation is expected as both MLP and TPBNN are designed to work with a fixed network topology. So, when the network topology changes, their performance suffers. 

\begin{figure}[h]
\centerline{\includegraphics[width=9cm]{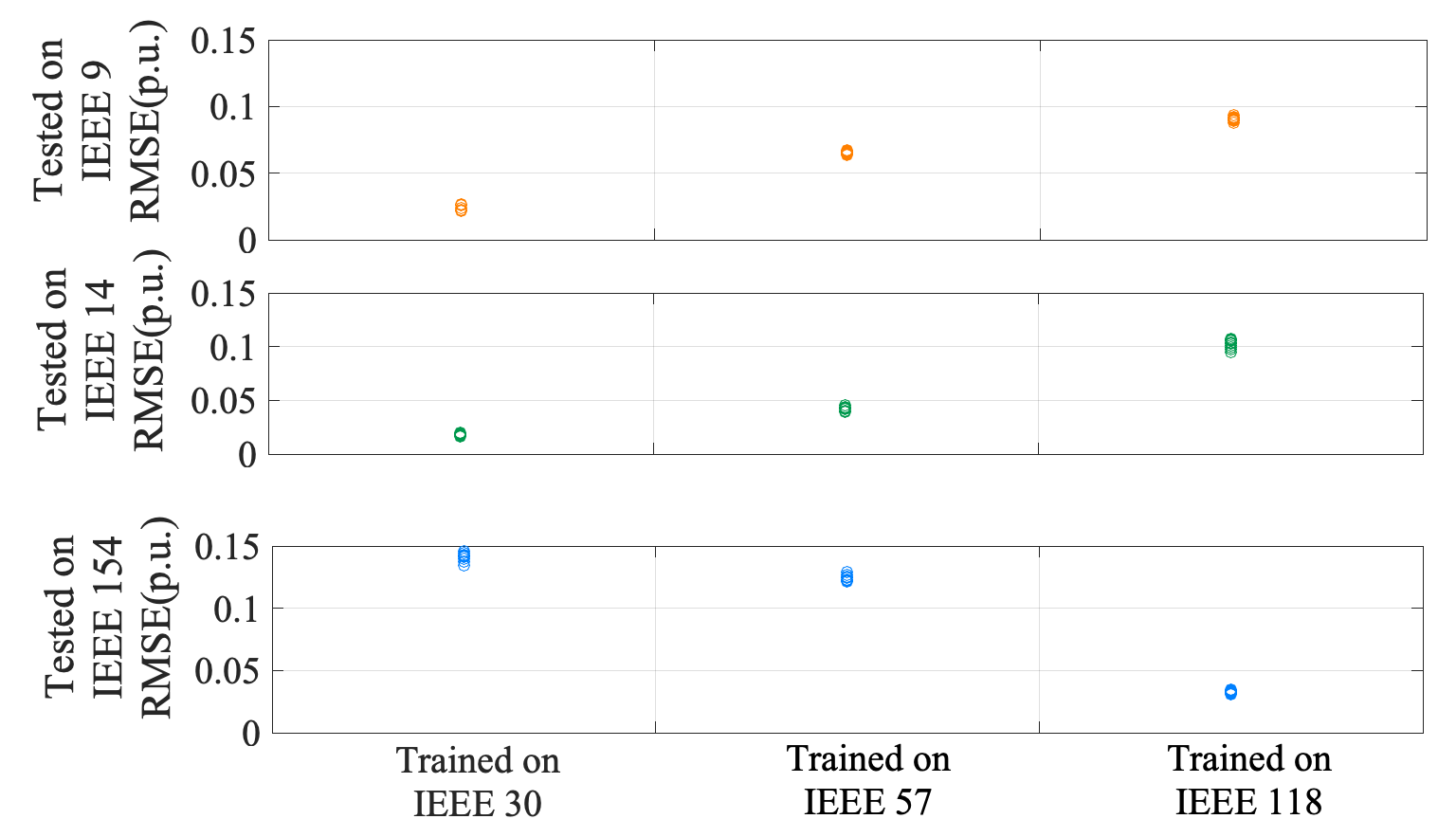}}
\caption{Rebuilt PF RMSE under different bus node configurations.}
\label{fig:new_cases}
\end{figure}

Additionally, we further show the generalization capability of our GAT models for new power networks where even the number of bus nodes are different 
from the training data sets. In this setting, we train different GAT models by using data sets under different base cases. In our experiment, three GAT models are trained with data sets generated from the IEEE 30-bus, IEEE 57-bus and IEEE 118-bus base cases, respectively. We then apply the three GAT models to different power networks that are generated from other three different base cases, i.e., IEEE 9-bus, 14-bus, and 154-bus cases, respectively. As shown in Figure~\ref{fig:new_cases}, the generalization errors are all small and stable (with a relative tight spread of their RMSE errors). The errors become even smaller when the test configurations are similar in size to the training configurations.

These experiments clearly demonstrate the good generalization capability of GAT over MLP and TPBNN.

\section{Conclusion}
\label{sec:diss}
\noindent
With the increasing uncertainties in modern power grids, it becomes difficult to acquire full power flow models for conducting various power flow analyses. Most of the existing works based on the ACPF formulation need a full power flow model to be useful. 
To address this issue, we propose a novel method based on graph attention networks to rebuild ACPF models 
by considering the realistic settings where the power system parameters can be inaccurate or even unavailable.
Experimental results show that our proposed GAT model has achieved better accuracy for changing power networks than two state-of-the-arts models, MLP [10] and TPBNN [21].
Moreover, different from MLP and TPBNN which are only applicable to a fixed network topology, our proposed model can generalize well to different power networks with varying topology configurations. Through our model, we can effectively build a mapping rule from the bus voltage to power injections, which can be then used for downstream tasks, such as as a regularizer for inverse prediction problems. One of our future directions is to combine the proposed model to an ACPF solver while considering its generalizability to different power networks.

\end{document}